\documentclass{article}
\usepackage{spconf,amsmath,graphicx}

\usepackage{subfig}
\usepackage{multirow}
\usepackage{booktabs}
\usepackage{array}
\newcolumntype{C}[1]{>{\centering\arraybackslash}p{#1}}
\RequirePackage[textfont=it,tableposition=top]{caption}

\newcommand{\bs}{\boldsymbol}
\newcommand{\tabincell}[2]{\begin{tabular}{@{}#1@{}}#2\end{tabular}}
\newenvironment{myalign}{%
    \setlength{\abovedisplayskip}{1.0ex}%
    \setlength{\belowdisplayskip}{1.0ex}%
    \align
  }%
  {\endalign}

\title{ResNeXt and Res2Net Structures for Speaker Verification}
\name{Tianyan Zhou, Yong Zhao, Jian Wu}
\address{Microsoft Corporation, USA \\
\normalsize \{tizhou, yonzhao, jianwu\}@microsoft.com}

\begin{document}
%
\maketitle
\begin{abstract}
The ResNet-based architecture has been widely adopted to extract speaker embeddings for text-independent speaker verification systems. By introducing the residual connections to the CNN and standardizing the residual blocks, the ResNet structure is capable of training deep networks to achieve highly competitive recognition performance. However, when the input feature space becomes more complicated, simply increasing the depth and width \footnote{width refers to the number of channels in a layer.} of the ResNet network may not fully realize its performance potential. In this paper, we present two extensions of the ResNet architecture, ResNeXt and Res2Net, for speaker verification. Originally proposed for image recognition, the ResNeXt and Res2Net introduce two more dimensions, cardinality and scale, in addition to depth and width, to improve the model's representation capacity. 
By increasing the scale dimension, the Res2Net model can represent multi-scale features with various granularities, which particularly facilitates speaker verification for short utterances. We evaluate our proposed systems on three speaker verification tasks. Experiments on the VoxCeleb test set demonstrated that the ResNeXt and Res2Net can significantly outperform the conventional ResNet model. The Res2Net model achieved superior performance by reducing the EER by 18.5\% relative. Experiments on the other two internal test sets of mismatched conditions further confirmed the generalization of the ResNeXt and Res2Net architectures against noisy environment and segment length variations.
\end{abstract}


\section{Introduction}
\label{sec:intro}
Speaker verification (SV) refers to the process of verifying a person's claimed identity using their enrolled recordings and test speech signals. It can be used to verify individuals for secure, frictionless user engagements in a wide range of solutions, from customer identity verification in call centers to contactless facility access. 
According to different application scenarios, SV can be categorized as text-dependent speaker verification (TD-SV) and text-independent speaker verification (TI-SV). For TD-SV, the spoken content of the test utterance and the enrollment utterance should be the same, whereas there is no constraint on the spoken content in the TI-SV system.

In the past few years, the architecture of SV systems has undergone a transformation from human designed subsystems \cite{dehak2010front, Prince2007Probabilistic, Villalba2011Towards, Garcia2011Analysis} to end-to-end frameworks. Many studies \cite{Variani2014Deep, heigold2016end, Snyder2017Deep, snyder2018x} have demonstrated the efficiency and capacity of various deep neural network structures. Most recently, ResNet-based \cite{he2016deep} convolutional neural networks (CNN) have been widely adopted as the backbone structure for TI-SV systems \cite{li2017deep, vox1, vox2, cai2018exploring, zhou2019cnn}. By introducing the residual connections to the CNN, the ResNet model is able to build very deep neural networks and achieve superior performance under challenging real-life conditions.

Typically, the ResNet network is customized in two essential dimensions, width and depth, to control the model's representation capacity accommodating the amount of training data. However, when the  input  feature  space  becomes  more  complicated and there exist various kinds of mismatches between the training and test conditions, simply increasing the depth and width of the ResNet network is prone to overfitting, thus impairing the generalization ability of the overall network.
Many approaches \cite{huang2017densely, xie2017aggregated, fisher2018dla, shang2019res2net} have been proposed with modified ResNet-based structures to maximize feature utilization and achieve better performance while keeping the model size manageable.

In this paper, we propose to incorporate two extensions of the ResNet structures, ResNeXt \cite{xie2017aggregated} and Res2Net \cite{shang2019res2net} into the speaker verification systems, given their encouraging performance on various image recognition tasks. The ResNeXt and Res2Net expose two more dimensions, cardinality and scale respectively, in addition to depth and width. The ResNeXt block replaces the residual block in the ResNet with a multi-branch transformation by introducing many groups of convolution in one layer. The Res2Net redesigns the residual block. It constructs hierarchical residual-like connections inside the residual block and assembles variable-size receptive fields within one layer. 
With the scale dimension, the Res2Net model can represent multi-scale features with various granularity, which facilitates speaker verification for very short utterances. 
We evaluate the proposed systems on several speaker verification tasks. Experiments on the VoxCeleb data set show that increasing cardinality is more efficient than going deeper or wider. Increasing the scale is even more powerful than increasing the other three dimensions. Experiments on the other two internal test sets confirmed the effectiveness of the proposed methods. We further truncated the original VoxCeleb test set into 2-4s segments. Experiments show the multi-scale feature representation ability in Res2Net could largely boost the performance for short utterances.

The rest of this paper is organized as follows: Section \ref{sec:cnn_embedding} describes the fundamental components of our ResNet-based speaker veriﬁcation baseline system. Section \ref{sec:residual_block} introduces the detailed implementation of ResNeXt block and Res2Net block in our system. Experimental results are discussed in Section \ref{sec:experiments}. In Section \ref{sec:conclusions}, we conclude this paper and present future work.

\section{ResNet-based Speaker Embedding}
\label{sec:cnn_embedding}
In this section, we describe four fundamental components in our end-to-end speaker verification system. First of all, we apply an utterance-level normalization to the input features. Then a ResNet is utilized to perceive feature patterns and generate frame-level speaker representations. In order to produce fixed length utterance-level speaker embedding from variable length audio inputs, an attentive pooling layer is added. Lastly, we select a  speaker-discriminative criterion to guide the training.

\subsection{Utterance-level mean normalization}
According to the experimental results in \cite{yong2020impro}, applying utterance-level mean normalization to input features can improve system robustness towards mismatched acoustic environment in test condition, while variance normalization does not help. Therefore, in our system, only mean normalization is used.

Given a sequence of input features $\{\bs{m}_1, \bs{m}_2, ..., \bs{m}_{T} \}$, extracted from one utterance, we first calculate the mean $\bs{\mu}$ along the temporal axis, and then normalize them as follows: 
\begin{myalign}
\bs{\mu} &= \frac{1}{T} \sum_{t=1}^{T} \bs{m}_t \\
\bs{\hat{m}}_t &=  \bs{m}_t - \bs{\mu}
\end{myalign}
where $\bs{\hat{m}}_t$ is the normalized feature; we then feed them into the following neural network.

\subsection{Network architecture}
In Table~\ref{tab:network}, we list the detailed configuration of our speaker embedding extractor. The backbone structure is ResNet with 17 convolutional layers. This network is based on the architecture in our previous work \cite{zhou2019cnn} with increased size and improved accuracy. It takes 80-dimensional log Mel filter banks as input features and generates speaker embedding with a dimension of 128. All convolutional layers are followed by a batch normalization layer and rectified linear units (ReLU) activation function. In order to preserve more frame-level speaker embeddings for attentive pooling, we set the stride to 1 along temporal axis in layer 'conv4' and 'conv5'. 

\begin{table}[htb]
	\caption{ResNet-based system configuration. Notation for convolutional layer: (channel, kernel size, stride).}
	\label{tab:network}
	\centering
	\begin{tabular}{|C{0.1\linewidth}|C{0.45\linewidth}|C{0.3\linewidth}|}
		\hline
		\textbf{Stage} & \textbf{Module} & \textbf{Output Size}\\
		\hline
		\tabincell{c}{input} & \tabincell{c}{mean normalization} & \tabincell{c}{$80 \times T \times 1$} \\
		\hline
		\tabincell{c}{conv1} & \tabincell{c}{$(64, 3\times 3, 2)$} & \tabincell{c}{$39 \times T/2 \times 64$} \\
		\hline
		\tabincell{c}{block1} & \tabincell{c}{$2\times\left[ \begin{array}{c} {64, 3\times3, 1}  \\ {64, 3\times3, 1} \end{array}\right]$} & \tabincell{c}{$39 \times T/2 \times 64$}\\
		\hline
		\tabincell{c}{conv2} & \tabincell{c}{$(128, 3\times 3, 2)$} & \tabincell{c}{$19 \times T/4 \times 128$} \\
		\hline
		\tabincell{c}{block2} & \tabincell{c}{$2\times\left[ \begin{array}{c} {64, 3\times3, 1}  \\ {64, 3\times3, 1} \end{array}\right]$} & \tabincell{c}{$19 \times T/4 \times 128$}\\
		\hline
		\tabincell{c}{conv3} & \tabincell{c}{$(256, 3\times 3, 2)$} & \tabincell{c}{$9 \times T/8 \times 256$} \\
		\hline
		\tabincell{c}{block3} & \tabincell{c}{$2\times\left[ \begin{array}{c} {128, 3\times3, 1}  \\ {128, 3\times3, 1} \end{array}\right]$} & \tabincell{c}{$9 \times T/8 \times 256$}\\
		\hline
		\tabincell{c}{conv4} & \tabincell{c}{$(256, 3\times 3, (2,1))$} & \tabincell{c}{$4 \times T/8 \times 256$} \\
		\hline
		\tabincell{c}{conv5} & \tabincell{c}{$(128, 3\times 3, (2,1))$} & \tabincell{c}{$1 \times T/8 \times 128$} \\
		\hline
		& \tabincell{c}{attentive pooling layer} & \tabincell{c}{$1 \times 128$}\\
		\hline
		& \tabincell{c}{classification layer} & \tabincell{c}{$128 \times 5994$}\\
		\hline
	\end{tabular}
\end{table}

\subsection{Multi-head attentive pooling}
This layer aggregates frame-level speaker embeddings and generates a more stable utterance-level speaker embedding. As reported in our previous work \cite{zhou2019cnn}, compared with simple temporal average pooling, attentive pooling has the potential to actively select speaker-discriminative frames. With multiple attention heads, we can further improve system performance by a large margin. In this work, we still choose the multi-head attentive pooling as our aggregation approach. In consideration of the tradeoff between performance and model complexity, the number of attention heads is set to 16. Implementation and other hyper parameters remain the same as the ones in \cite{zhou2019cnn}.

\subsection{Training criterion}
In order to directly compare the representation power of different model structures, we didn't employ any modified softmax criteria or other discriminative loss functions. During the training phase, after extracting the utterance-level speaker embedding, we feed it into a classification layer and use the naive softmax loss.

\section{Residual block}
\label{sec:residual_block}
ResNet is a highly modularized architecture. Modules of same topology are stacked together to construct networks with different representation power. In this case, less hyper parameters are exposed for tuning, potentially retaining the robustness and generalization of the ResNet structure. 

In ResNet, depth and width are two major dimensions to tune the model capacity. Typically, a model with more parameters possesses a stronger representation power. However, in practice, we usually find directly going deeper or wider is not efficient. Besides, it could easily suffer from overfitting due to the huge amount of parameters. In order to ease this problem and effectively improve ResNet's representation capacity, two structures are proposed by image recognition community, called ResNeXt \cite{xie2017aggregated} and Res2Net \cite{shang2019res2net}. Basically, they redesign the residual blocks in conventional ResNet and expose two new dimensions to change model capacity.

In this section, we describe the design of residual blocks in ResNeXt and Res2Net, and their implementations in our system.

\subsection{ResNeXt block}
\begin{figure}[htb]

	\centering
	\includegraphics[trim= 20mm 5mm 130mm 55mm, clip, width=1\linewidth]{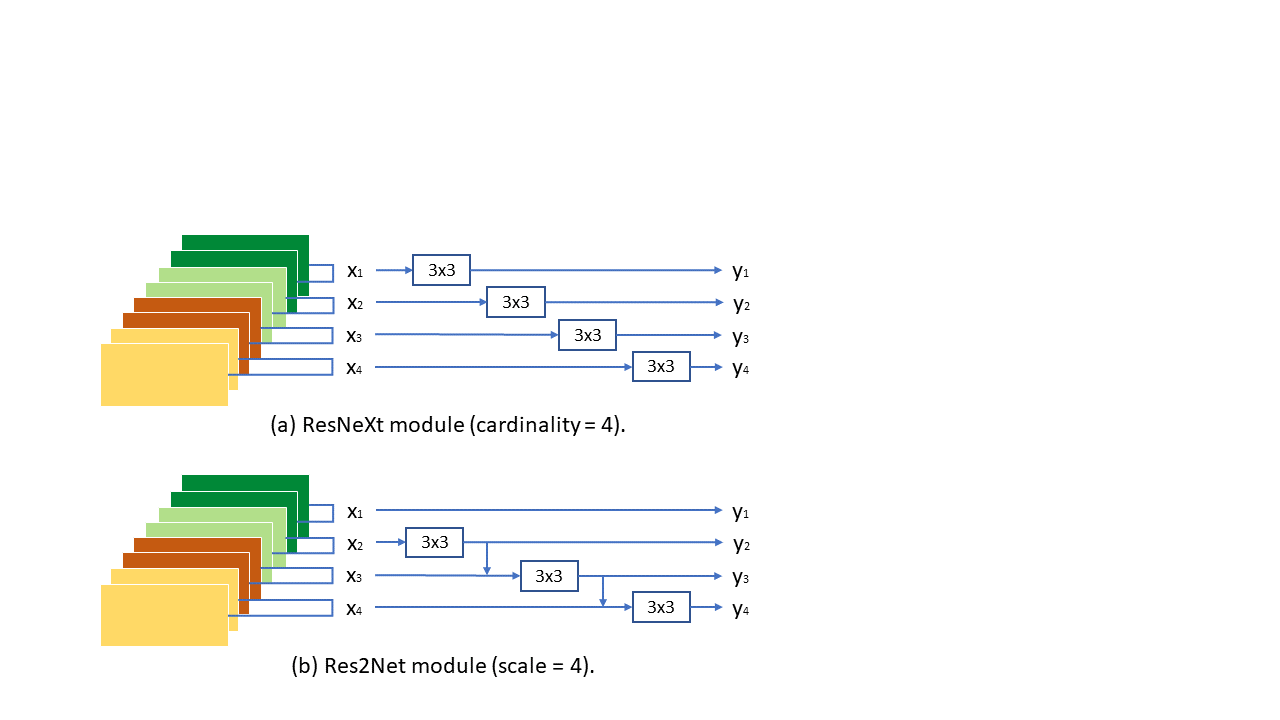} 
	\caption{Detailed designs inside ResNeXt and Res2Net blocks..}
	\label{fig:modules}
\end{figure}

A ResNeXt \cite{xie2017aggregated} block follows the split-transform-merge strategy but performs a set of transformations instead of a single one in the ResNet block. As illustrated in Figure~\ref{fig:modules}(a), after splitting the input channels into several smaller groups (4 groups as an example), 4 ways of transformations are then applied separately on those groups. The outputs from all groups are merged through a $1\times1$ convolutional layer. The size of the set of transformations is named as \textit{cardinality}. \cite{xie2017aggregated} showed increasing cardinality is more effective and efficient than going wider or deeper. 

\begin{figure}[htb]

	\centering
	\includegraphics[trim= 20mm 65mm 130mm 30mm, clip, width=1\linewidth]{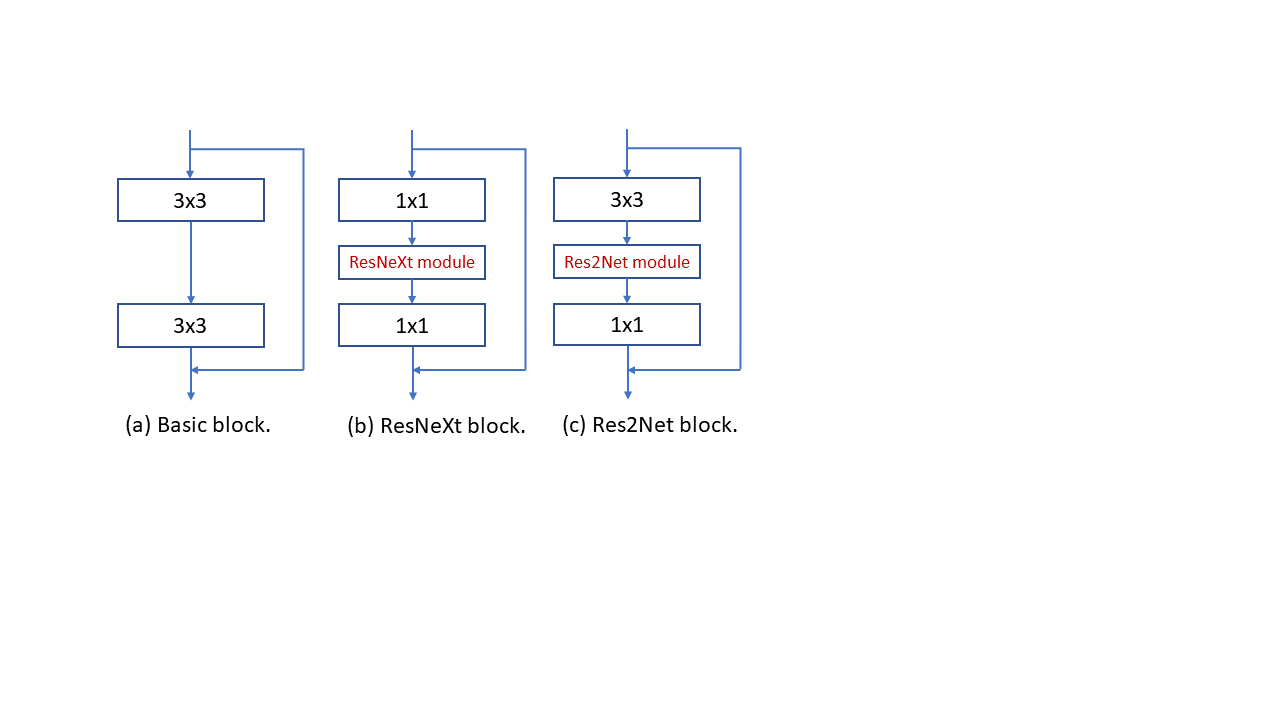} 
	\caption{Three types of residual blocks.}
	\label{fig:blocks}
\end{figure}

Since ResNeXt block shares the same topology, we can directly use it to replace the residual block in conventional ResNet and easily stack them to construct the desired structures. For our network showed previously, there are 6 residual blocks in total and all of them share the topology in Figure~\ref{fig:blocks}(a), which has a depth of 2. \cite{xie2017aggregated} has demonstrated applying aggregating transformations to Figure~\ref{fig:blocks}(a) only makes it wider, which is not the nontrivial topology we want. As a result, we start with bottleneck block (depth = 3) showed in Figure~\ref{fig:blocks}(b), and implement the multi-branch transformations at the second layer.


\subsection{Res2Net block}
A Res2Net \cite{shang2019res2net} block aims at improving model's multi-scale representation ability by increasing the number of available receptive fields within one layer. When stacking several blocks together, we would obtain feature representations from various granularities due to combination effects. Figure~\ref{fig:modules}(b) illustrates the design of Res2Net module. After going through the first $3\times3$ layer, feature maps are evenly sliced into $s$ subsets, denoted by $\{x_1, x_2, ..., x_{s}\}$. Each subset is then fed into a $3\times3$ convolution (denoted by $\mathbf{K_i}$), except for $x_1$. Starting from $x_3$, each output of $\mathbf{K}_{i-1}$ is added with $x_i$ before going through $\mathbf{K}_i$. This hierarchical residual-like connection further increases the possible receptive fields within one layer, resulting in multiple feature scales. The whole process can be formulated as 

\begin{equation}
\label{eq:angleloss}
y_i=\begin{cases}
  x_i, & i=1; \\
  \mathbf{K}_i(x_i), & i=2; \\
  \mathbf{K}_i(x_i+y_{i-1})), & i=3,4, ... , s \\
  \end{cases}
\end{equation} 
where $\{y_1, y_2, ..., y_{s}\}$ is the output of this module, they are concatenated and fed into the following $1\times1$ convolutional layer to maintain the channel size of this residual block. In \cite{shang2019res2net}, $s$ is named as \textit{scale}. They proved increasing this new dimension is more effective than other dimensions (width, depth and cardinality).

In order to modify the basic block in our baseline ResNet structure to a Res2Net block, we keep the first $3\times3$ layer and replace the second one with a Res2Net module, as shown in Figure~\ref{fig:blocks}(c).

\begin{table*}[htb]
	\caption{Evaluation results of different models on three test conditions, i.e., VoxCeleb1-test, VoxCeleb1-E, and VoxCeleb1-H. EER and minDCF (p-target=0.01) are reported. Notation for model: (w: base width; c: cardinality; s: scale). Preserved indicates that the model keeps the same complexity as the baseline ResNet model; Increased indicates higher complexity.}
	\label{tab:vox-result}
	\centering
	\begin{tabular}{c l r c c c c c c}
		\toprule
		\multirow{2}{*}{\textbf{Complexity}} & 
		\multirow{2}{*}{\textbf{Model}} & 
		\multirow{2}{*}{\textbf{\# Params}} & 
		\multicolumn{2}{c}{\textbf{VoxCeleb1-test}} & 
		\multicolumn{2}{c}{\textbf{VoxCeleb1-E}} & 
		\multicolumn{2}{c}{\textbf{VoxCeleb1-H}} \\
		\cmidrule{4-9} & & & \textbf{EER(\%)} & \textbf{minDCF}  & \textbf{EER(\%)} & \textbf{minDCF} & \textbf{EER(\%)} & \textbf{minDCF} \\ 
    	\midrule
		\multirow{7}{*}{Preserved} & ResNet & 5.2M & 1.78 & 0.1734 & 1.76 & 0.1931 & 3.07 & 0.2950\\
		& ResNeXt-40w4c & 5.4M & 1.69 & 0.1427 & 1.71 & 0.1921 & 3.00 & 0.2869 \\
		& ResNeXt-26w8c & 5.3M & 1.77 & 0.1723 & 1.72 & 0.1849 & 2.97 & 0.2853\\
		& ResNeXt-12w32c & 5.9M & 1.69 & 0.1545 & 1.66 & 0.1796 & 2.90 & 0.2743\\
		& Res2Net-48w2s & 5.5M & 1.75 & 0.1833 & 1.70 & 0.1815 & 2.98 & 0.2831\\
		& Res2Net-26w4s & 5.6M & 1.68 & 0.1522 & 1.63 & 0.1789 & 2.90 & 0.2878\\
		& Res2Net-14w8s & 5.6M & 1.60 & 0.1775 & 1.60 & 0.1840 & 2.83 & 0.2797\\
		\midrule
		\multirow{5}{*}{Increased} & ResNet-deeper & 7.4M & 1.77 & 0.1835 & 1.76 & 0.1998 & 3.22 & 0.3041 \\
		& ResNet-wider & 7.7M & 1.81 & 0.2025 & 1.76 & 0.1895 & 3.09 & 0.2988 \\
	    & ResNeXt-20w32c & 10.2M & 1.61 & 0.1445 & 1.57 & 0.1739 & 2.78 & 0.2674 \\
		& Res2Net-26w6s & 7.5M & 1.51 & 0.1544 & 1.54 & 0.1754 & 2.75 & 0.2740\\
		& Res2Net-26w8s & 9.3M & \textbf{1.45} & \textbf{0.1471} & \textbf{1.47} & \textbf{0.1692} & \textbf{2.72} & \textbf{0.2717} \\
		\bottomrule
	\end{tabular}
\end{table*}

\section{Experiments}
\label{sec:experiments}

\subsection{Dataset Description}
We evaluate the proposed approach on the VoxCeleb corpus and the other two Microsoft internal test sets. 

\textbf{VoxCeleb}: VoxCeleb \cite{vox1, vox2} is a public text-independent dataset with real-life conversational speech collected in unconstrained conditions. It contains diverse acoustic environments and short-term utterances, making it a more challenging task than telephone recordings or clean speech. All the  models are trained on the VoxCeleb2-dev part of 5994 speakers. 
To improve the robustness of the model, we  apply four augmentation techniques (babble, music, noise, and reverb) to increase the diversity of the training data, following the Kaldi recipe in \cite{snyder2018x}. 
We evaluate our models on the VoxCeleb1 test set composed of 40 identities. For completeness, we also evaluate two more test conditions: the extended VoxCeleb1-E, which uses the entire VoxCeleb1 set (1251 identities), and hard VoxCeleb1-H list which contains speakers with same gender and nationality. 
 
\textbf{MS-SV test set}: This is an internal test set collected for the experimental purpose of text-independent speaker verification (SV). Each participant is enrolled by reading a short paragraph through a close microphone. Then we recorded their daily interactions in several Microsoft meeting rooms through far field microphones and post-processed as SV test set. The set contains around 150 speakers. the test utterances ranges from 2-15s (about 4s in average). The test set is an extended version of the MS-SV test set used in \cite{yong2020impro} with more speakers and improved front-end processing. 

\textbf{Cortana test set}: This is the  internal test set recorded for text-dependent speaker verification. We collect a set of utterances that start with the `Cortana' or `Hey Cortana' wake-up phrases from 183 speakers. Each participant is asked to read a list of such utterances under various noise environments with different recording distances. Four enrollment utterances are recorded in clean condition, 1 meter away from the microphone. We cut out test segments using a CTC-based keyword detector. The CTC endpoint is shifted backward 1.5s and forward 2s, resulting in average length of 3s for processed test segments. The keyword itself has an average duration of 0.7s.

\subsection{Experiments on VoxCeleb1 test sets}
\subsubsection{Regular test set}
Table~\ref{tab:vox-result} shows the performance of different residual blocks on VoxCeleb1 test set. $w$ refers to the base width of each $3\times3$ convolution inside ResNeXt or Res2Net module. We can adjust $w$ to preserve model complexity while increasing cardinality or scale. 

With preserved complexity, ResNet baseline produces EER of 1.78\%, 1.76\%, and 3.07\% for VoxCeleb1-test, VoxCeleb1-E and VoxCeleb1-H, respectively. ResNeXt and Res2Net achieve 1.69\% and 1.60\% EER on the VoxCeleb1-test task, which outperforms the ResNet model. 

When we increase the model complexity by 50\%, the ResNet model (`ResNet-deeper' and `ResNet-wider') does not yield further gain. In contrast, the Res2Net model `Res2Net-26w6s' yields further improvements. When we increase the model complexity by around 80\%, the model 'Res2Net-26w8s' achieves the best result on all three test conditions. Specifically, it produces 1.45\% EER on the VoxCeleb1-test set, outperforming the ResNet baseline by 18.5\% relative.
Moreover, these experiments basically show that introducing cardinality and scale is more effective than simply going deeper or wider. In addition, increasing scale is more effective than increasing cardinality. 

Furthermore, it is observed that training accuracy of 'ResNeXt-20w32c', 'Res2Net-26w8s' and 'ResNet' are 99.11\%, 99.07\%, and 98.41\%, respectively. This implies the gain come from stronger representation power, other than regularization.

\subsubsection{Truncated test set}
The utterances in the VoxCeleb1-test set have an average duration of 8s. 
We derive a truncated test set of short segments by keep the trial pairs and truncating the test utterances into segments of 2s, 3s and, 4s, respectively. These test sets are used to evaluate the multi-scale feature representation ability of the proposed models. 
Table~\ref{tab:vox1_test_short} shows that the performance degrades considerably on short segments. However, ResNeXt and Res2Net still outperform ResNet in this challenging condition. Particularly, if we compare Res2Net with ResNet baseline, the EER reduction for 2s, 3s, 4s is 17.6\%, 19.0\%, and 13.7\%, respectively. The Res2Net model shows significant advantage for short utterances.

\begin{table}[htb]	\caption{Evaluation results with different model structures. EER(\%) is reported on VoxCeleb1-test truncated set.}
	\label{tab:vox1_test_short}
	\centering
	\label{tab:ms-sv}
	\centering
    \begin{tabular}{l c c c c}
    	\toprule
    	\textbf{Model}  &   \textbf{2s} & \textbf{3s} & \textbf{4s} & \textbf{Regular} \\ \midrule        
        ResNet  &   6.77 & 3.78  &	2.49  & 1.78 \\ \midrule
        ResNeXt-20w32c & 	6.26 & 3.35  &	2.35  & 1.61 \\ \midrule
        Res2Net-26w8s & 	5.58 & 3.06  &	2.15  & 1.45 \\ \midrule
    \end{tabular}  	
\end{table}

\subsection{Experiments on text-independent MS-SV test set}
In this section, we evaluate the model trained with VoxCeleb data on the text-independent MS-SV test set to examine the generalization of the proposed models, as shown in Table~\ref{tab:ms_sv}. The enrollment utterances are 20s long in average. We split the trials into three groups in terms of test segment duration (1-2s, 2-4s, and $>$4s). Res2Net shows better generalization over ResNeXt, with 8.3\% (EER: 5.28\% $\rightarrow$ 4.84\%) relative reduction in the overall EER over the ResNet baseline.

\begin{table}[htb]
	\caption{Evaluation results with different model structures. EER(\%) is reported on MS-SV test set.}
	\label{tab:ms_sv}
	\centering
    \begin{tabular}{l c c c c}
	\toprule
	\textbf{Model}  & \textbf{1-2s} & \textbf{2-4s} & \textbf{$>$4s} & \textbf{Overall} \\ \midrule
    ResNet  & 7.66	& 3.67	& 2.11 & 5.28 \\ \midrule
    ResNeXt-20w32c & 7.42	& 3.63	& 2.04 & 5.11 \\ \midrule
    Res2Net-26w8s	& 6.77	& 3.32	& 2.03 & 4.84 \\ \bottomrule
    \end{tabular}  	
\end{table}

The results of different duration intervals are also consistent with what we have observed on VoxCeleb1 test set. A breakdown of EER by test utterance length shows that Res2Net model is capable of producing more powerful speaker representations from very short speech segments. The relative improvement over ResNet is 11.6\%, 9.5\%, and 3.8\% for interval 1-2s, 2-4s, and $>$4s, respectively. 

\subsection{Experiments on text-dependent Cortana test set}

We also conduct some experiments on the text-dependent Cortana test set. For convenience, we still use the aforementioned models trained on VoxCeleb data. As a result, the spoken content between the training and test data is mismatched, but the text between the enrollment and test utterance is matched.

Table~\ref{tab:cortana_sv} shows the overall EER and a breakdown of the performance in terms of noise conditions and distance to microphones. The Res2Net outperforms the ResNet by 4.5\% (EER: 4.40\% $\rightarrow$ 4.20\%) relative in the overall EER. Specifically, the Res2Net improves over the ResNet by 11.7\%, 11.2\% in quiet and TV conditions, respectively.
ResNeXt performs worse than the ResNet. We conjecture that the ResNeXt model is not as robust as Res2Net in the mismatched conditions.

\begin{table}[htb]
\setlength{\tabcolsep}{3.5pt}
	\caption{Evaluation results with different model structures. EER(\%) is reported on Cortana test set}
	\label{tab:cortana_sv}
	\centering
    \begin{tabular}{l c c c c c c c}
    	\toprule
    	\multirow{ 2}{*}{\textbf{Model}} & 
    	\multicolumn{2}{c}{\textbf{Condition}} & 
    	\multicolumn{3}{c}{\textbf{Distance}} & 
    	\multirow{ 2}{*}{\textbf{Total}} \\  
    	\cmidrule{2-6}
    	 & \textbf{Quiet} & \textbf{TV}  & \textbf{1m} & \textbf{3m} &	\textbf{5m} &  \\ \midrule
        ResNet  &	3.26 &	5.91 &	3.11	& 4.62	& 5.51 & 4.40 \\ \midrule
        ResNeXt-20w32c &   3.31 &  5.80 & 	2.95	& 5.11	& 5.26 & 4.60 \\ \midrule
        Res2Net-26w8s &	2.88 &	5.25  &	2.71	& 4.21	& 4.85 & 4.20 \\ \bottomrule
    \end{tabular} 
\end{table}

\begin{figure*}[t]
    \centering
    \subfloat[]{\includegraphics[trim= 190mm 15mm 185mm 20mm, clip, width=0.5\textwidth]{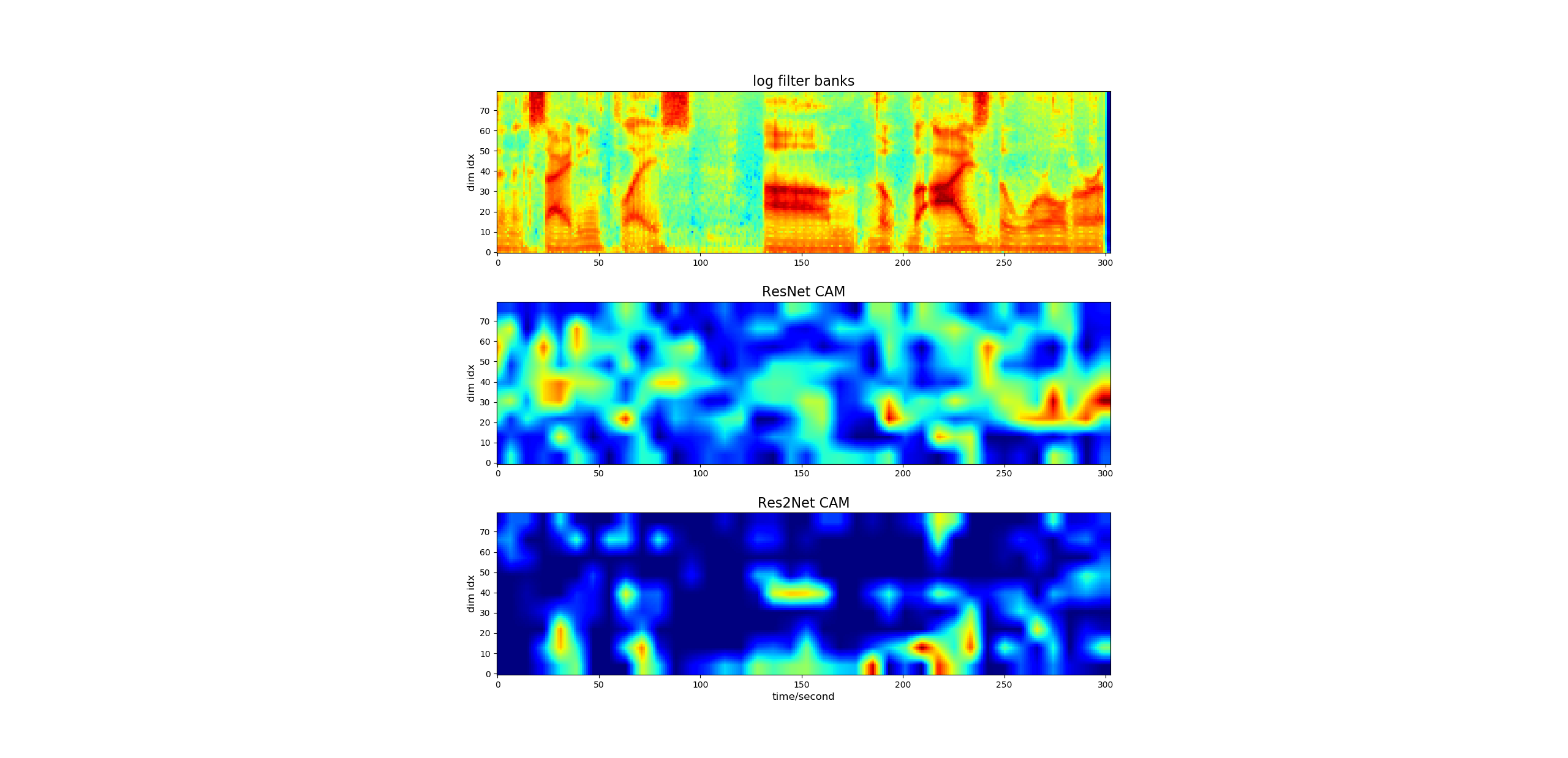}} 
    \subfloat[]{\includegraphics[trim= 190mm 15mm 185mm 20mm, clip, width=0.5\textwidth]{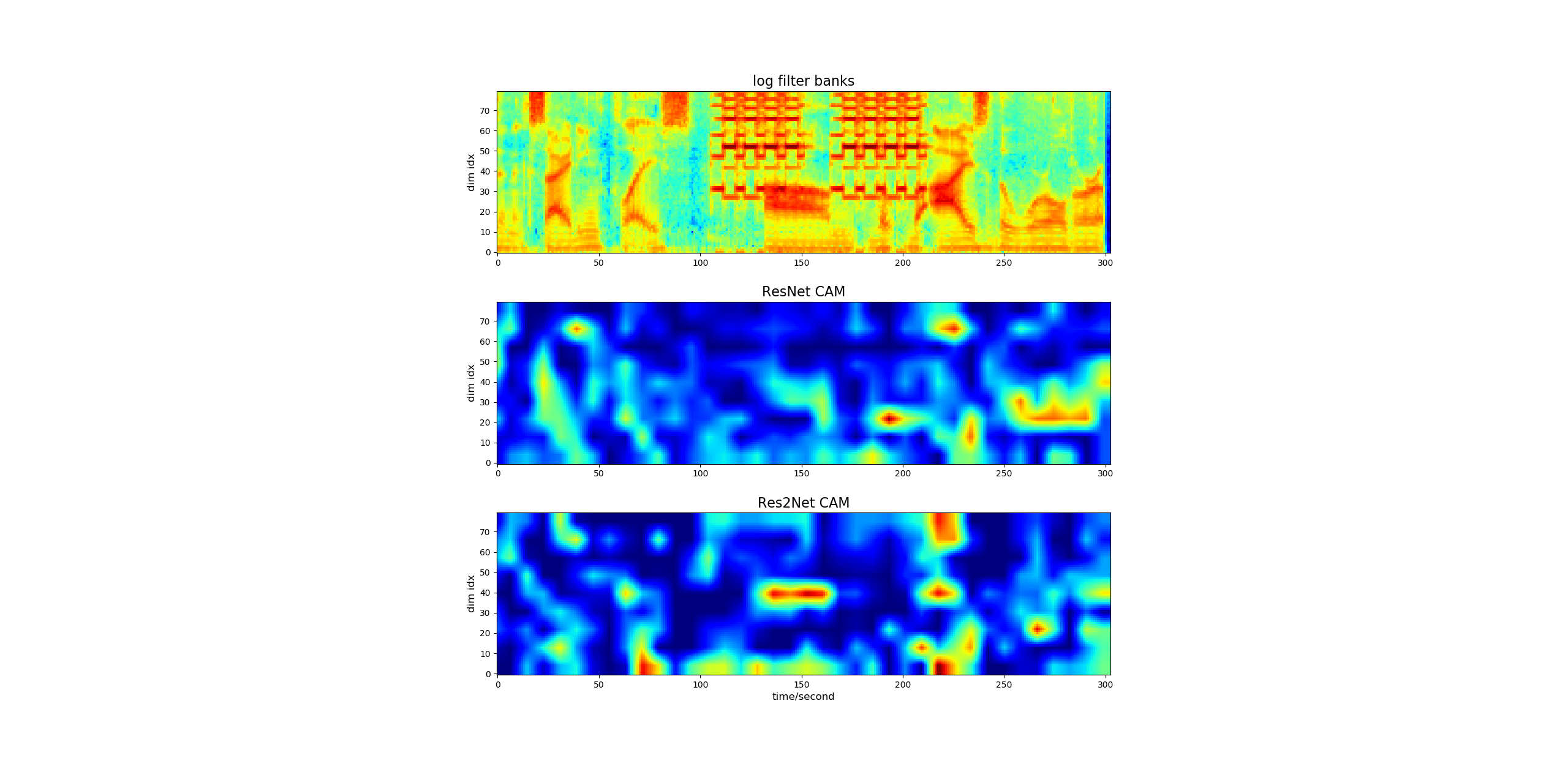}} 
    \caption{(a). ResNet CAM and Res2Net CAM of utterance "id00018-r377L5cuPOw-00144. (b). ResNet CAM and Res2Net CAM of utterance "id00018-r377L5cuPOw-00144-noise".}
    \label{fig:cam}
\end{figure*}

\subsection{Visualization}
Inspired by the class activation mapping (CAM) analysis in \cite{shang2019res2net}, we intend to provide some intuitive insights into the capacity of Res2Net model on speech data. We picked an utterance in clean condition as well as its simulated version with telephone ring noise. We feed two utterances into ResNet and Res2Net separately and plot the CAM using Grad-CAM \cite{selvaraju2017grad}.

As shown in Figure~\ref{fig:cam}, for clean speech, Res2Net CAM is more concentrated than ResNet CAM. The latter spreads over most parts of input features, including non-speech part. For the noise-corrupted speech,  Res2Net CAM shows less changes on the activation map. In addition, compared with the Res2Net CAM in clean condition, the remaining speech part is further emphasized. This example may help us understand the capacity of Res2Net structure. It might have the potential to extract more invariant feature representations and recognize speakers in adverse environments.

\section{Conclusions}
\label{sec:conclusions}

In this paper, we investigate the effectiveness of ResNeXt and Res2Net architectures on the speaker verification task. Experimental results on VoxCeleb test set demonstrate their strong representation powers. Both networks outperform the ResNet baseline, and Res2Net exhibits even stronger capacity. It keeps improving system performance for short utterances and mismatched scenarios. Experiments on two internal test sets provide us more confidence on the generalization of Res2Net models. Our future work includes incorporating ResNeXt and Res2Net structures with discriminative loss or additional phonetic information.

\bibliographystyle{IEEEbib}
\bibliography{refs}

\end{document}